\documentclass[10pt,pre,twocolumn,bibnotes,nofootinbib,superscriptaddress,byrevtex,showpacs,letterpaper]{revtex4}
\usepackage{graphics}
\usepackage{amssymb}
\usepackage{bm}
\usepackage{color}
\usepackage[normalem]{ulem}

\newcommand{\be}{\begin{equation}}
\newcommand{\ee}{\end{equation}}
\renewcommand{\emph}{\textit}
\newcommand{\ra}{\rightarrow}
\newcommand{\bW}{\bar{W}}
\newcommand{\bM}{\bar{M}}
\newcommand{\bV}{\bar{V}}
\newcommand{\bw}{\bar{w}}

\newcommand{\lex}{\left\langle}
\newcommand{\rex}{\right\rangle}

\renewcommand{\emph}{\textit}
\newcommand{\rmd}{\mathrm{d}}
\newcommand{\rme}{\mathrm{e}}
\newcommand{\rmi}{\mathrm{i}}

\begin{document}

\title{Anomalous fluctuation relations}

\author{H. Touchette}
\email{ht@maths.qmul.ac.uk}
\affiliation{School of Mathematical Sciences, Queen Mary University of London, London E1 4NS, UK}

\author{E. G. D. Cohen}
\affiliation{The Rockefeller University, 1230 York Avenue, New York, New York 10021, USA}

\date{\today}

\begin{abstract}
We complement and extend our work on fluctuation relations arising in nonequilibrium systems in steady states driven by L\'evy noise [Phys.\ Rev.\ E 76, 020101(R) (2006)]. As a concrete example, we consider a particle subjected to a drag force and a L\'evy white noise with tail index $\mu\in (0,2]$, and calculate the probability distribution of the work done on the particle by the drag force, as well as the probability distribution of the work dissipated by the dragged particle in a nonequilibrium steady state. For $0<\mu<2$, both distributions satisfy what we call an \textit{anomalous fluctuation relation}, characterized by positive and negative fluctuations that asymptotically have the same probability. For $\mu=2$, by contrast, the work and dissipated work distributions satisfy the known \emph{conventional} and \emph{extended fluctuation relations}, respectively, which are both characterized by positive fluctuations that are exponentially more probable than negative fluctuations. The difference between these different  fluctuation relations is discussed in the context of large deviation theory. Experiments that could probe or reveal anomalous fluctuation relations are also discussed.
\end{abstract}

\pacs{05.40.-a, 02.50.-r, 05.70.-a}

\maketitle

\section{Introduction} 
\label{secintro}

Many works on nonequilibrium systems have been devoted recently to the study of fluctuations around out-of-equilibrium steady states. The basis of these studies is typically the following: for a given system driven in a nonequilibrium steady state, one considers an observable $A_\tau$ integrated over a time $\tau$, such as the work done by a force acting on the system during a time $\tau$, and proceeds to calculate its probability distribution $P(A_\tau)$. Interestingly, what has been found for many different systems and many observables $A_\tau$ is that the positive fluctuations of $A_\tau$ are exponentially more probable than its negative fluctuations, in the sense that
\be
\frac{P(A_\tau=\tau a)}{P(A_\tau=-\tau a)}=\rme^{c \tau a+o(\tau)},
\label{eqfr1}
\ee
where $c$ is some constant that does not depend on $\tau$, and $o(\tau)$ is a sublinear correction term in $\tau$, which may be zero if the probability ratio is exactly exponential in $\tau$. Examples of observables for which Eq.~(\ref{eqfr1}) is observed include the entropy production of chaotic deterministic systems \cite{gallavotti1995,gallavotti1995a,gallavotti1998a} and stochastic Markov systems \cite{kurchan1998,lebowitz1999,maes1999}, as well as other entropy- and work-like quantities defined in the context of particles moving in fluids \cite{wang2002,zon2003a}, electrical circuits \cite{zon2004a,garnier2005}, granular media \cite{aumaitre2001,feitosa2004,puglisi2005,visco2005}, and turbulent fluids \cite{ciliberto1998,ciliberto2004}, among other systems.

In early studies of these systems, it was thought that the result expressed by Eq.~(\ref{eqfr1}) might be a general law of nonequilibrium fluctuations, but it is now known that this is not the case. Some observables, such as the heat absorbed by driven Brownian particles \cite{farago2002,zon2003,zon2004} and the current of the zero-range process \cite{harris2006,rakos2008}, satisfy a more general relation of the form
\be
\frac{P(A_\tau=\tau a)}{P(A_\tau=-\tau a)}=\rme^{\tau f(a)+o(\tau)},
\label{eqfr2}
\ee
where $f(a)$ is in general a nonlinear function of $a$, which does not depend on $\tau$. The difference between Eqs.~(\ref{eqfr1}) and (\ref{eqfr2}) serves, following the work of van Zon and Cohen \cite{zon2003,zon2003a}, as a basis for classifying nonequilibrium fluctuations. Observables that comply with Eq.~(\ref{eqfr1}) are said to satisfy a \textit{conventional fluctuation relation}, whereas those satisfying the more general Eq.~(\ref{eqfr2}) are said to satisfy an \textit{extended fluctuation relation}. 

Our goal in this paper is to revisit this classification of nonequilibrium fluctuations. Based on the fact that the existence of a conventional or extended fluctuation relation is essentially equivalent to the existence of a probability distribution having a large deviation form \cite{touchette2009}, we have constructed in \cite{touchette2007} a model of a driven nonequilibrium system for which the mechanical work $W_\tau$ done over a time $\tau$ by the driving force satisfies neither the conventional nor the extended fluctuation relation because the probability distribution $P(W_\tau)$ of $W_\tau$ fails to have the form of a large deviation probability. The probability distribution $P(W_\tau)$ is nevertheless well defined, and can be used to define the probability ratio $P(W_\tau)/P(-W_\tau)$, which has in this case a power-law rather than an exponential form in $\tau$. We have called this property of the probability ratio an \textit{anomalous fluctuation relation}, following a terminology used in studies of L\'evy-type noise, and have proposed some experiments with which one could physically ``realize'' or ``test'' this type of fluctuation relation.

Here we complete our study of anomalous fluctuation relations initiated in \cite{touchette2007} by discussing in more detail the fluctuations of $W_\tau$ for the model studied in that paper, and by considering the fluctuations of an additional observable, called dissipated work. These two points are the subject of Secs.~\ref{secwork} and \ref{secdissw}, respectively. We also discuss in Sec.~\ref{secfrldt} the difference between conventional and extended fluctuation relations, on the one hand, and the anomalous fluctuation relation, on the other hand, from the general point of view of large deviation theory. The main result discussed in that section follows the observation that fluctuation relations of the conventional and extended types are equivalent to having probabilities of the large deviation kind, and that other types of fluctuation relations must arise whenever the probability distribution of an observable does not have the large deviation form. The anomalous fluctuation relation that we discuss here is but one example of fluctuation relations, which arise by replacing Gaussian white noise as the source of noise in Langevin equations by L\'evy white noise or, more generally, by replacing noise sources having finite moments by noise sources having infinite moments.

The relationship between fluctuation relations and large deviation theory was noted in the original derivations of the fluctuation theorem for the entropy production \cite{gallavotti1995,gallavotti1995a,gallavotti1998a} (see also \cite{gallavotti2008}), and is explicitly discussed in our previous paper \cite{touchette2007}, as well as in a recent review paper written by one of us \cite{touchette2009}. Here we continue these discussions by studying some conditions under which the anomalous fluctuation relation is expected to arise. We end the paper in Sec.~\ref{secdisc} with various remarks related to the physical interpretation of our results, the nature of L\'evy noise, as well as future work aimed at extending and experimentally verifying our results.

\section{Model}
\label{secmodel}

We consider in this paper a Brownian particle subjected to three different forces: a linear restoring force arising from a particle-confining harmonic potential moving at a constant speed, a friction force, and a random force or noise. The Langevin equation modelling the effect of these forces on the Brownian particle can be found in the work of van Zon and Cohen \cite{zon2003a}, which is itself based on the experimental work of Wang \textit{et al.}~\cite{wang2002}. Here we study the overdamped version of that equation, given by
\be
\alpha\dot x=-\kappa [x(t)-vt]+\xi(t). 
\label{eqle1}
\ee 
In this equation, $x(t)$ denotes the position (in the laboratory frame) of the Brownian particle at time $t$, $v$ is the velocity with which the harmonic potential moves, $\alpha$ is the friction coefficient, $\kappa$ is the strength of the harmonic potential, and $\xi(t)$ is the random force. 

In previous studies of the model defined above, as in most studies of Langevin equations, the random force $\xi(t)$ is assumed to be a Gaussian white noise characterized by its zero mean $\langle\xi(t)\rangle=0$ and its autocorrelation function $\langle\xi(t)\xi(t')\rangle=\Gamma\delta(t-t')$, where $\Gamma$ is the noise power. We depart from this assumption here by taking $\xi(t)$ to be a \emph{symmetric L\'evy white noise}, defined by the following characteristic functional:
\be
G_\xi[k]=\int\mathcal{D}[\xi]\ P[\xi]\ e^{\rmi\int k(t)\xi(t) \rmd t}=\exp\left(-b \int |k(t)|^\mu\, \rmd t\right),
\label{eqcfn1}
\ee
with $b>0$ and $0<\mu\leq 2$ \cite{gnedenko1954,uchaikin1999,samoradnitsky2000}. The first integral in this expression is the path integral defining the characteristic function of the noise process $\xi(t)$, whereas the second expression on the right-hand side is the expression that defines $\xi(t)$ as being a L\'evy white noise with strength $b$ and index $\mu$; that is, an uncorrelated noise distributed according to a symmetric L\'evy distribution with scale parameter $b$ and tail index $\mu$.\footnote{In our previous work, we used $\alpha$ to denote the index of the L\'evy noise. Here we use $\mu$ for this index, since we use $\alpha$ to denote the friction coefficient.} The case of Gaussian white noise considered in \cite{zon2003a} is recovered from this characteristic function by choosing $\mu=2$, in which case $\xi(t)=0$ and $\langle\xi(t)\xi(t')\rangle=\Gamma\delta(t-t')$ with $\Gamma=2b$ as the noise power.

It is important to note that the use of L\'evy noise in Langevin equations is often seen as problematic, since the mean of $\xi(t)$ diverges for $\mu\in (0,1)$, and its noise power $\Gamma$ diverges for all $\mu\in (0,2)$ \cite{gnedenko1954,uchaikin1999,samoradnitsky2000}. The fact is, however, that these divergences do not lead to any physical pathologies; they are merely the sign that the concept of mean and noise power (viz., variance) are ill-defined or inapplicable for L\'evy noise. This point has been discussed in the literature; see, e.g., \cite{uchaikin1999}. In particular, the fact that $\xi(t)$ has an infinite mean for $\mu\in (0,1)$ does not imply physically that the energy supplied to the Brownian particle by the random force $\xi(t)$ is infinite. The increments of $\xi(t)$ are necessarily always finite, so that the energy supplied by $\xi(t)$ is also always finite.

As for the divergence of the power of $\xi(t)$ for $\mu\in(0,2)$, it is true that it implies that there can be no fluctuation-dissipation relation relating the friction coefficient with the noise power. However, if we view the friction and the noise as being independent, that is, as arising physically from different physical mechanisms, then there is no need for a fluctuation-dissipation relation between the friction and the noise. This situation is physically possible: one can imagine, for example, that the friction force in the Langevin equation arises from a solid-solid contact, while the noise is imposed externally, say, using a computer-generated L\'evy noise. This situation will be discussed more concretely in Sec.~\ref{secdisc}. 

For now we will leave these considerations aside to focus on the nonequilibrium steady state generated by the Langevin equation defined in Eq.~(\ref{eqle1}), and to study the fluctuations of the work done on the Brownian particle in such a state by the moving potential. This is done in the next section. The fluctuations of the dissipated work are studied in Sec.~\ref{secdissw}.

\section{Work fluctuations}
\label{secwork}

The total work done on the Brownian particle described by Eq.~(\ref{eqle1}) is the sum of two contributions: the mechanical work done on the particle by the random force $\xi(t)$ \cite{farago2002}, which is explicitly considered here as an independent \textit{external} force, and the mechanical work done on the particle by the moving harmonic potential \cite{zon2003a}. We study in this section the latter quantity which is given by the integral 
\be
W_\tau=-\kappa v\int_0^{\tau} [x(t)-vt]\, \rmd t
\ee 
for a given time interval $[0,\tau$]. Thus $W_\tau$ is the mechanical work done on the Brownian particle by the harmonic potential over a time $\tau$. For convenience, we rewrite this quantity as
\be
W_\tau=-\kappa v \int_0^{\tau} y(t)\, \rmd t
\label{eqw1}
\ee
in terms of the position $y(t)=x(t)-vt$ of the particle in the frame of the moving potential (comoving frame) \cite{zon2003a}. In terms of $y(t)$, the Langevin equation of Eq.~(\ref{eqle1}) reads
\be
\dot y(t)=-\frac{1}{\tau_r}y(t)-v+\frac{1}{\alpha}\xi(t),
\label{eqle2}
\ee
where we have defined $\tau_r=\alpha/\kappa$, the characteristic relaxation time of the particle in the potential.

Our aim in the next subsections is to calculate the probability distribution of $W_\tau$ for various values of $\mu$, and to discuss the properties of this distribution in the light of fluctuation relations. 

\subsection{General distribution}

The calculation of the distribution $P(W_\tau)$ is done in three steps. First, note that the characteristic function of $W_\tau$ can be obtained from the characteristic functional $G_y[k]$ of $y(t)$, defined as 
\be
G_y[k]=\left\langle \exp\left(\rmi\int_0^\infty k(s)\, y(s)\, \rmd s\right)\right\rangle,
\ee 
by choosing the test function
\be
k(s)=
\left\{
\begin{array}{lll}
-q\kappa v, & & 0\leq s\leq \tau\\
0, & & s>\tau.
\end{array}
\right.
\label{eqk1}
\ee
From the definition of the work, given in Eq.~(\ref{eqw1}), we indeed have
\be
G_{W_\tau}(q)=\langle \rme^{\rmi q W_\tau}\rangle=\lex\exp\left(-\rmi q\kappa v\int_0^\tau y(t)\, \rmd t\right)\rex=G_y[k]
\ee

In the second step, we express the characteristic function of $y(t)$ in terms of the characteristic function of the noise process $\xi(t)$. The Langevin equation for $y(t)$, shown in (\ref{eqle2}), is linear in $y(t)$, so we can use for this purpose a general result of C\'aceres and Budini~\cite{caceres1997,caceres1999}, given by
\be
G_y[k]=\exp\left(\rmi r_0 y_0-\rmi v\int_0^\infty r(l)\, \rmd l\right) G_\xi[r/\alpha],
\label{eqcfy1}
\ee
where
\be
r(l)=\int_l^\infty e^{(l-s)/\tau_r}\, k(s)\, \rmd s,
\label{eqr1}
\ee
$r_0=r(0)$, $y_0=y(0)=x(0)$, and $G_\xi$ is the characteristic function of the noise $\xi(t)$. Substituting the expression of $G_\xi$, given in Eq.~(\ref{eqcfn1}), we then obtain
\be
G_{W_\tau}(q)=\exp\left(\rmi r_0y_0-\rmi v\int_0^\infty r(l)\, \rmd l-\frac{b}{\alpha^\mu} \int_0^\infty |r(l)|^\mu\, \rmd l \right),
\label{eqcfw1}
\ee
with $r(l)$ given by Eq.~(\ref{eqr1}) and $k(s)$ given by Eq.~(\ref{eqk1}). Equations~(\ref{eqcfy1}) and (\ref{eqcfw1}) assume that the initial condition $y_0$ is a constant. If $y_0$ is a random variable, then $G_y[k]$ must be averaged over the distribution of $y_0$ to obtain the proper characteristic function of the complete process $y(t)$ with its random initial condition. Here we shall assume that $y_0=0$.

The third and final step leading to $P(W_\tau)$ consists in evaluating the integrals involved in Eqs.~(\ref{eqr1}) and (\ref{eqcfw1}). The integral in Eq.~(\ref{eqr1}) defining $r(l)$ has, with Eq.~(\ref{eqk1}), the solution
\be
r(l)=
\left\{
\begin{array}{lll}
q v \alpha (\rme^{(l-\tau)/\tau_r}-1), & & 0\leq l \leq \tau\\
0, & & l>\tau.
\end{array}
\right.
\label{eqr2}
\ee
From this result, we compute the first integral on the right-hand side of Eq.~(\ref{eqcfw1}):
\be
\int_0^\infty r(l)\, \rmd l=v \alpha q [\tau_r(1-\rme^{-\tau/\tau_r})-\tau].
\ee
As for the second integral appearing on the right-hand side of Eq.~(\ref{eqcfw1}), which involves $\mu$, it cannot be solved analytically for all $\mu\in(0,2]$, to the best of our knowledge, although it is clear that it leads to a term proportional to $|q|^\mu$ in the exponential of $G_{W_\tau}(q)$, since $r(l)$ is proportional to $q$ according to Eq.~(\ref{eqr2}). As a result, we obtain
\be
G_{W_\tau}(q)=\rme^{\rmi Mq-V|q|^\mu},
\label{eqcfw2}
\ee
where
\be
M=v^2\alpha [\tau-\tau_r(1-\rme^{-\tau/\tau_r})],
\label{eqm1}
\ee
and
\be
V=bv^\mu \int_0^\tau |\rme^{-(\tau-l)/\tau_r}-1|^\mu\, \rmd l.
\label{eqv1}
\ee

\begin{figure}[t]
\includegraphics{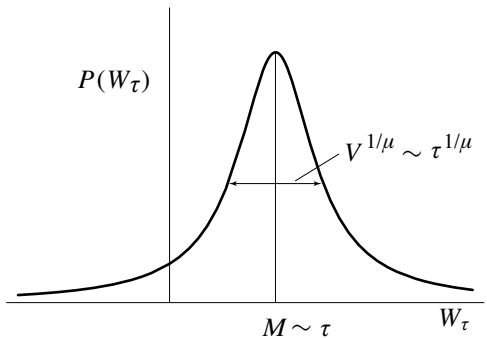}
\caption{Sketch of the distribution $P(W_\tau)$ of the total (extensive) work. The mode $M$ of the distribution is proportional to $\tau$ while the width, which is proportional to $V^{1/\mu}$, scales like $\tau^{1/\mu}$.}
\label{figscaling1}
\end{figure}

The characteristic function of Eq.~(\ref{eqcfw2}) is a central result of this paper, which shows, following the theory of L\'evy distributions \cite{gnedenko1954,uchaikin1999,samoradnitsky2000}, that the distribution $P(W_\tau)$ of the work $W_\tau$ is a symmetric L\'evy distribution having the following properties:

(i) $P(W_\tau)$ is symmetric and centered around $M$, so that $M$ represents the most probable value or \emph{mode} of $W_\tau$; see Fig.~\ref{figscaling1}.

(ii) For $1<\mu\leq 2$, the integral
\be
\langle W_\tau\rangle=\int_{-\infty}^\infty w\, P(W_\tau=w)\, \rmd w
\ee
defining the mean of $W_\tau$ exists, so that $M$ also represents the mean of $W_\tau$. This applies only for this range of $\mu$ values; for the complementary range $0<\mu\leq 1$, the integral above does not converge, so that $M$ cannot be interpreted as the mean. Thus $M$ represents the mode for all $\mu\in (0,2]$, but represents the mean only for $\mu\in (1,2]$.

(iii) The parameter $V$ is related to the width of the distribution $P(W_\tau)$: the larger $V$ is, the wider $P(W_\tau)$ is. For $\mu=2$, in particular, $V$ is half the variance of $W_\tau$, so the width of $P(W_\tau)$, taken as the standard deviation of $W_\tau$, is proportional to $V^{1/2}$. For $0<\mu<2$, the variance does not exist even though $V$ is finite. In this case, one can still relate $V$ to the width of $P(W_\tau)$ by calculating moments of $W_\tau$ of order smaller than $2$. In particular, it can be proved \cite{west1982} that
\be
\langle |W_\tau-M|^\beta \rangle = a V^{\beta/\mu},
\label{eqvsim1}
\ee
where $0<\beta<\mu$ and $a$ is some positive constant. The average $\langle |W_\tau-M|^\beta \rangle$ is called the \emph{fractional moment of order} $\beta$ \cite{west1982}. As a particular case of Eq.~(\ref{eqvsim1}), we have for $\beta=1$:
\be
\langle |W_\tau-M|\rangle\sim a V^{1/\mu}
\ee
when $\mu>1$.

(iv) $M$ and $V$ are directly proportional to $\tau$ in the limit of large $\tau$, that is, $M\sim \tau$ and $V\sim\tau$ as $\tau\ra\infty$. This follows from Eqs.(\ref{eqm1}) and (\ref{eqv1}). 

(v) For $0<\mu<2$, $P(W_\tau)$ has power-law tails that decay according to
\be
P(W_\tau=w)\sim \frac{\mu V}{ |w-M|^{\mu+1}}
\label{eqpw1}
\ee
as $|w|\ra\infty$ \cite{montroll1987}. This property is responsible for the divergent mean observed for $\mu\in(0,1]$ and the divergent variance observed for $\mu\in (0,2)$. Whenever, $P(W_\tau)$ has this property, we say that the tails of $P(W_\tau)$ decay as a power-law with \emph{exponent} $\mu$.

We will be interested in the next sections to study the fluctuations of $W_\tau$ in the long-time or asymptotic limit where $\tau$ is much larger than the relaxation time $\tau_r$. To properly define this limit, we must note two properties of $W_\tau$. First, because $M$ is proportional to $\tau$, according to the property (iv) above, the mode of $P(W_\tau)$ escapes to infinity as $\tau\ra\infty$. Second, because $V$ is also proportional to $\tau$, the width of $P(W_\tau)$ also grows indefinitely with $\tau$. Thus $P(W_\tau)$ gets flatter as $\tau\ra\infty$ while its mode moves to infinity. 

To eliminate these diverging properties of $P(W_\tau)$, we consider the \emph{intensive} or \emph{scaled work}, defined by $\bW_\tau=W_\tau/M$, as the random variable of interest rather than total work $W_\tau$ which is extensive with $\tau$.\footnote{The scaled work $W_\tau/M$ is equal to $W_\tau/\langle W_\tau\rangle$ when the average work $\langle W_\tau\rangle$ exists.} The characteristic function of $\bW_\tau$ is related to the characteristic function of $W_\tau$ by a simple rescaling:
\be
G_{\bW_\tau}(k)=G_{W_\tau}(k/M)=\rme^{\rmi k\bM -\bV |k|^\mu}.
\label{eqcfbw1}
\ee 
With this change of variables, it is easily seen that the mode of $\bW_\tau$ is $\bM=1$, while its ``width'' is given by $\bV=V/M^\mu$. Therefore, the probability distribution of $\bW_\tau$ is now centered at 1, and is such that $\bV\sim\tau^{1-\mu}$ as $\tau\ra\infty$, since $V\sim\tau$ and $M\sim\tau$ in that limit.

Note that we could have considered $W_\tau/\tau$ as the scaled work rather than $W_\tau/M$. The only difference between these two definitions is the value of the mode $\bM$: for the scaled work defined by $W_\tau/M$, $\bM$ is always equal to 1, whereas for $\bW_\tau/\tau$, $\bM$ is a positive constant which may be different than 1. Note also that rescaling the extensive work $W_\tau$ by $\tau$ or any factor proportional to $\tau$ is the only way of centering the distribution of $W_\tau$ to a constant. For, if one divides $W_\tau$ by a factor smaller than $\tau$, then the mode of the resulting random variable will still grow with $\tau$, whereas if one divides $W_\tau$ by a factor greater than $\tau$, then the distribution of the resulting variable will become symmetric. In the latter case, all the information about asymmetry of the work fluctuations is discarded.

With this in mind, we now turn to studying the distribution of $\bW_\tau$ for different values of $\mu$ in the range $[0,2]$. Four cases of fluctuations arising from four different regimes of L\'evy noise will be considered. The first is the Gaussian noise regime that leads to a conventional fluctuation relation for $\bW_\tau$. The three others are proper L\'evy noise regimes that lead to the anomalous fluctuation relation for $\bW_\tau$.

\subsection{Gaussian noise: $\mu=2$}
\label{secgausswork}

The distribution of the scaled work $\bW_\tau=W_\tau/M$ that results from Gaussian white noise (i.e., $\mu=2$) was found by van Zon and Cohen \cite{zon2003a}. We quickly repeat their main results here, since they serve as our point of departure for defining the anomalous fluctuation relation. What is important to note is that the distribution $P(\bW_\tau)$ for $\mu=2$ can be put in the form
\be
P(\bW_\tau=\bw)=\rme^{-\tau I(\bw)+o(\tau)},
\label{eqldw1}
\ee 
where
\be
I(\bw)=\frac{(\bw-1)^2}{2}.
\label{eqrf2}
\ee
The parabolic form of the function $I(\bw)$, which is called the \emph{rate function} \cite{touchette2009}, obviously leads to a Gaussian distribution for $P(\bW_\tau)$. For simplicity, we shall drop in the remaining the $o(\tau)$ term in probability distributions so as to write
\be
P(\bW_\tau=\bw)\sim\rme^{-\tau I(\bw)}.
\label{eqldpw2}
\ee
The approximation sign ``$\sim$'' means, following Eq.~(\ref{eqldw1}), that $P(\bW_\tau=\bw)$ decays, to a first degree of approximation, exponentially with $\tau$. This property, which is commonly referred to as the \emph{large deviation property} or \emph{large deviation principle} \cite{touchette2009}, plays an important role in fluctuation relations, as it directly implies that
\be
\frac{P(\bW_\tau=\bw)}{P(\bW_\tau=-\bw)}\sim\rme^{\tau f(\bw)},
\label{eqldtfr1}
\ee
where
\be
f(\bw)=I(-\bw)-I(\bw).
\ee
What is special about the Gaussian noise case is that the \emph{fluctuation function} $f(\bw)$ happens to be linear in $\bw$, so that
\be
\frac{P(\bW_\tau=\bw)}{P(\bW_\tau=-\bw)}\sim\rme^{\tau \bw}.
\label{eqcftw1}
\ee
This result is the signature of the conventional fluctuation relation, as defined in \cite{zon2003,zon2003a} and Eq.~(\ref{eqfr1}) of Sec.~\ref{secintro}. Thus, under Gaussian noise, the mean work $\bW_\tau$ is said to satisfy a conventional fluctuation relation.\footnote{To be more precise, we should say that $\bW_\tau$ satisfies an \emph{asymptotic} conventional fluctuation relation, since the approximation above is only valid in the limit of large times $\tau$. All the results obtained in this paper are derived in this limit, so the attribute ``asymptotic'' will be omitted.}

\subsection{Cauchy noise: $\mu=1$}

\begin{figure*}[t]
\resizebox{\textwidth}{!}{\includegraphics{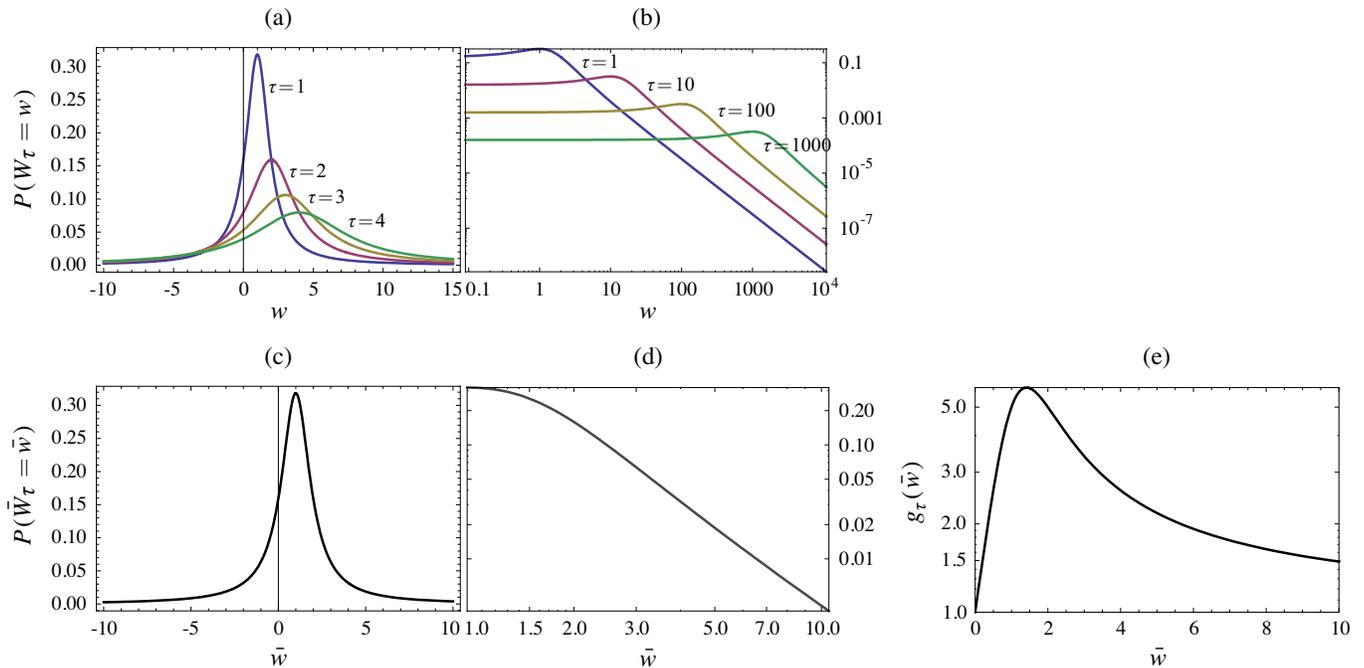}}
\caption{Work fluctuations for Cauchy noise, $\mu=1$. (a) Distribution $P(W_\tau=w)$ of the total (extensive) work $W_\tau$. (b) Log-log plot of $P(W_\tau=w)$ for positive $w$. (c) Distribution $P(\bW_\tau=\bw)$ of the scaled work $\bW_\tau$. (d) Log-log plot of $P(\bW_\tau=\bw)$ for positive $\bw$. The power-law tails of $P(W_\tau)$ and $P(\bW_\tau)$ give rise to straight lines with slope $-2$ in the log-log plots. (e) Fluctuation relation: Log-linear plot of $g_\tau(\bw)=P(\bW_\tau=\bw)/P(\bW_\tau=-\bw)$. Note that $P(\bW_\tau=\bw)$, and consequently $g_\tau(\bw)$, become invariant with $\tau$ for large $\tau$. This is the main feature of the Cauchy case. Units: $\tau_r=v=b=1$.}
\label{figa1}
\end{figure*}

The Fourier transform of the general characteristic function that we have derived for $\bW_\tau$ can be computed analytically for only two values of $\mu$. The first value is $\mu=2$, which we have just considered and which leads to Gaussian fluctuations of $\bW_\tau$. The second value is $\mu=1$ and leads to a so-called \emph{Cauchy distribution} having the form
\be
P(W_\tau=w)=\frac{1}{\pi}\frac{V}{(w-M)^2+V^2}
\label{eqpwa1}
\ee
for the total work and
\be
P(\bW_\tau=\bw)=\frac{1}{\pi}\frac{\bV}{(\bw-\bM)^2+\bV^2}
\label{eqpbwa1}
\ee
for the scaled work. These distributions were already discussed in our previous paper \cite{touchette2007}. What is important to note about $P(W_\tau)$ is that its mode $M$ and its width parameter $V$ are both proportional to $\tau$, which implies, as mentioned before, that this distribution moves to the right and flattens as $\tau$ increases. This is illustrated in Figs.~\ref{figa1}(a) and \ref{figa1}(b). The power-law tails of $P(W_\tau)$ predicted by the expression in Eq.~(\ref{eqpw1}) are clearly seen in the log-log plot of Fig.~\ref{figa1}(b) as straight lines.

The distribution $P(\bW_\tau)$ of the scaled work retains the power-law tails of $P(W_\tau)$, as is obvious from Eqs.~(\ref{eqpwa1}) and (\ref{eqpbwa1}) as well as from Figs.~\ref{figa1}(c) and \ref{figa1}(d), but does not have the translation and flattening behavior that we have noted for $P(W_\tau)$. In fact, one exceptional property of $P(\bW_\tau)$ for $\mu=1$ is that it becomes time-independent in the limit of large $\tau$ because $\bV=V/M=O(1)$ in $\tau$; see Figs.~\ref{figa1}(c) and \ref{figa1}(d). This property directly implies that the ratio $P(\bW_\tau=\bw)/P(\bW_\tau=-\bw)$ is also time independent. The precise form of this ratio, which we denote by $g_\tau(\bw)$ from now on, is obtained from Eq.~(\ref{eqpbwa1}):
\be
g_\tau(\bw)=\frac{P(\bW_\tau=\bw)}{P(\bW_\tau=-\bw)}\sim\frac{(-\bw-1)^2+1}{(\bw-1)^2+1},
\label{eqff1}
\ee
and is plotted on a log-linear scale in Fig.~\ref{figa1}(e). In this plot, $g_\tau(\bw)$ has a maximum located at $\bw=\sqrt{2}$, and
\be
\lim_{\bw\ra\infty} g_\tau(\bw)=1.
\label{eqlim1}
\ee

This last limit has no equivalent in the Gaussian case, and is a direct consequence of the fact that both tails of $P(\bW_\tau)$ decay as a power-law  with the same exponent $\mu=1$. In the Gaussian case, both tails of $P(\bW_\tau)$ decay exponentially, but because $P(\bW_\tau)$ is not centered at 0, the positive fluctuations of $\bW_\tau$ are exponentially more probable than the negative fluctuations. In the case of Cauchy noise, the difference between $P(\bW_\tau=\bw)$ and $P(\bW_\tau=-\bw)$ is so weak that the ratio $g_\tau(\bw)$ of these two probabilities goes to 1 in the limit of very large fluctuations. This implies, concretely, that large positive fluctuations are asymptotically just as likely to be observed as negative fluctuations of equal magnitude, as was already noted in \cite{touchette2007}.

We shall see next that this property of Cauchy fluctuations carries over to all other values of $\mu$ in the range $(0,2)$. For this reason, we follow our previous work \cite{touchette2007} and define a new class of fluctuation relations based on this property. Given the random variable $\bW_\tau$ and its distribution $P(\bW_\tau)$, we say that $\bW_\tau$ satisfies an \emph{anomalous fluctuation relation} if i) $P(\bW_\tau)$ has power-law tails; and ii) the limit (\ref{eqlim1}) is satisfied. The term ``anomalous'' follows the terminology used in studies of L\'evy-type noise, which refer, for example, to ``anomalous diffusion'' or ``anomalous transport'' as diffusion or transport processes driven by L\'evy noise or noises akin to L\'evy noise (see, e.g., \cite{klages2008}). In this context, the term ``anomalous'' is used not in the sense of ``abnormal'', but in the sense of ``anomalous'' with respect to the  ``normal'' behavior obtained with Gaussian noise.

Note that there is a further property of Gaussian fluctuations that we lose with Cauchy noise, namely, that the fluctuations of $\bW_\tau$ do not decrease as $\tau\ra\infty$. This again is a consequence of the time-independence of $P(\bW_\tau)$ obtained with Cauchy noise. The practical consequence of this difference is that, in the case of Cauchy noise, one does not need to accumulate a large number of samples of $\bW_\tau$ to observe deviations in the value of this quantity. For Gaussian noise, an exponential number (in $\tau$) of such samples is needed to observe deviations of $\bW_\tau$ from its mean. But for Cauchy noise any small samples of $\bW_\tau$ will reveal that this quantity is fluctuating no matter how large $\tau$ is.

\subsection{Upper L\'evy regime: $1<\mu<2$}

\begin{figure*}[t]
\resizebox{\textwidth}{!}{\includegraphics{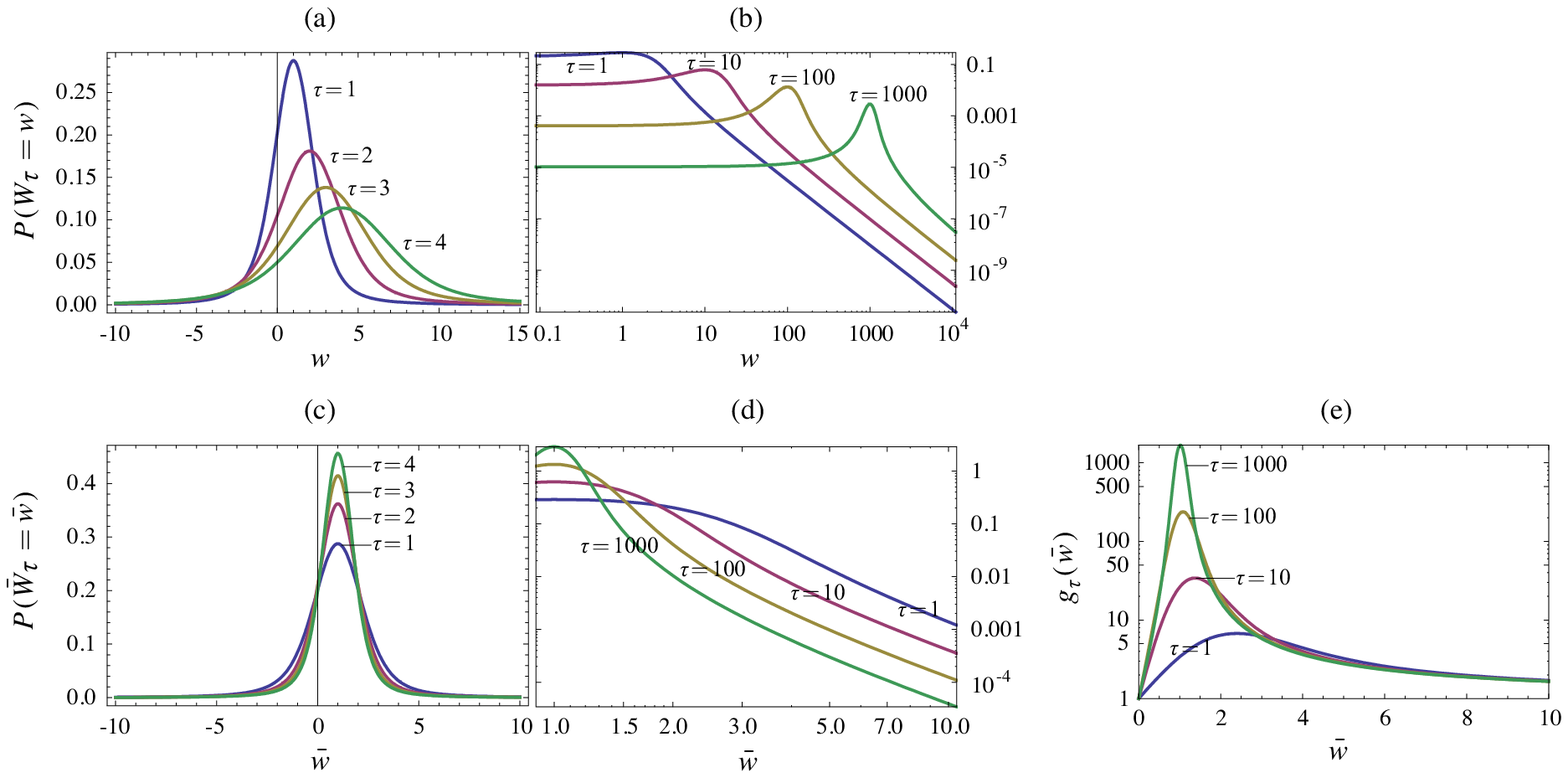}}
\caption{Work fluctuations for $\mu=1.5$ (upper L\'evy regime). (a) Distribution $P(W_\tau=w)$ of the total (extensive) work $W_\tau$. (b) Log-log plot of $P(W_\tau=w)$ for positive $w$. (c) Distribution $P(\bW_\tau=\bw)$ of the scaled work $\bW_\tau$. (d) Log-log plot of $P(\bW_\tau=\bw)$ for positive $\bw$. The power-law tails of $P(W_\tau)$ and $P(\bW_\tau)$ give rise to straight lines with slope $-\mu-1$ in the log-log plots. (e) Fluctuation relation: Log-linear plot of $g_\tau(\bw)=P(\bW_\tau=\bw)/P(\bW_\tau=-\bw)$. Note that $g_\tau(\bw)\ra 1$ as $\bw\ra\infty$ for all $\tau$. Units: $\tau_r=v=b=1$.}
\label{figa15}
\end{figure*}

Two different regimes of fluctuations arise when considering values of $\mu$ different from $1$ and $2$: a regime of ``weaker-than-Cauchy'' fluctuations corresponding to $\mu\in (1,2)$, and a regime of ``stronger-than-Cauchy'' fluctuations corresponding to $\mu\in(0,1)$. We discuss in this subsection the former regime, which we call the \emph{upper L\'evy regime}.

The plots shown in Fig.~\ref{figa15} illustrate the behavior of $P(W_\tau)$ and $P(\bW_\tau)$ for the case $\mu=1.5$, which is representative of all the values $\mu\in(1,2)$ in the upper L\'evy regime. These plots were obtained by numerically calculating the inverse Fourier transform of the characteristic function $G_{W_\tau}(q)$, found in Eq.~(\ref{eqcfw2}), using the ``Stable'' Mathematica package \cite{rimmer2005}. As in the Cauchy case, we see here that both $P(W_\tau)$ and $P(\bW_\tau)$ have power-law tails, and that the mode of $P(W_\tau)$ increases with $\tau$, whereas the mode of $P(\bW_\tau)$ is fixed at $1$. However, contrary to the Cauchy case, the distribution $P(\bW_\tau)$ of the scaled work for $\mu\in(1,2)$ is not invariant with $\tau$, but becomes more and more concentrated around the value $\bM=1$ as $\tau\ra\infty$; see Figs.~\ref{figa15}(c) and \ref{figa15}(d). This arises because typical fluctuations of $W_\tau$ increase less rapidly than $\tau$ in this range of $\mu$, which implies that the typical fluctuations for $\bW_\tau$ must decrease with $\tau$. This concentration of $P(\bW_\tau)$ around its mode implies a Law of Large Numbers for $\bW_\tau$, in the sense that $\bW_\tau\ra 1$ with probability 1 as $\tau\ra\infty$. This Law of Large Numbers also holds for Gaussian noise, and means concretely that measurements of $\bW_\tau$ are most likely to be close to 1 as one considers longer and longer integration times $\tau$.\footnote{A random variable for which the Law of Large Numbers holds is also said to be ``self-averaging''.} 

For large but finite $\tau$, fluctuations can still be observed, and a measure of how positive fluctuations are more likely to be observed than negative fluctuations is provided by the ratio $g_\tau(\bw)$. This ratio is plotted in Fig.~\ref{figa15}(e). As for the Cauchy case, we see here that positive fluctuations of the mean work are more likely to be observed than negative fluctuations of equal magnitude, since $g_\tau(\bw)>1$ for $\bw>0$, and that $g_\tau(\bw)\ra 1$ as $\bw\ra\infty$, meaning that the difference in probabilities for positive and negative fluctuations becomes negligible for large fluctuations. Because of this, and the fact that $P(\bW_\tau)$ has power-law tails, we conclude that $\bW_\tau$ satisfies an anomalous fluctuation relation for all $\mu\in (1,2)$, as in the Cauchy case. Unlike this case, however, the maximum of $g_\tau(\bw)$ increases with $\tau$ when $\mu\in (1,2)$, and moves towards the value $\bw=1$ as a result of the Law of Large Numbers. This behavior of the maximum of $g_\tau(\bw)$ is specific to $\mu\in (1,2)$ and thus serves as a signature of the upper L\'evy fluctuation regime. In the Cauchy case, by comparison, the fluctuations are stronger and lead to a time-independent $g_\tau(\bw)$ whereas, in the Gaussian case, the fluctuations are considerably weaker and lead to a ratio $g_\tau(\bw)$ which is exponential in $\tau$.

\subsection{Lower L\'evy regime: $0<\mu<1$}

\begin{figure*}[t]
\resizebox{\textwidth}{!}{\includegraphics{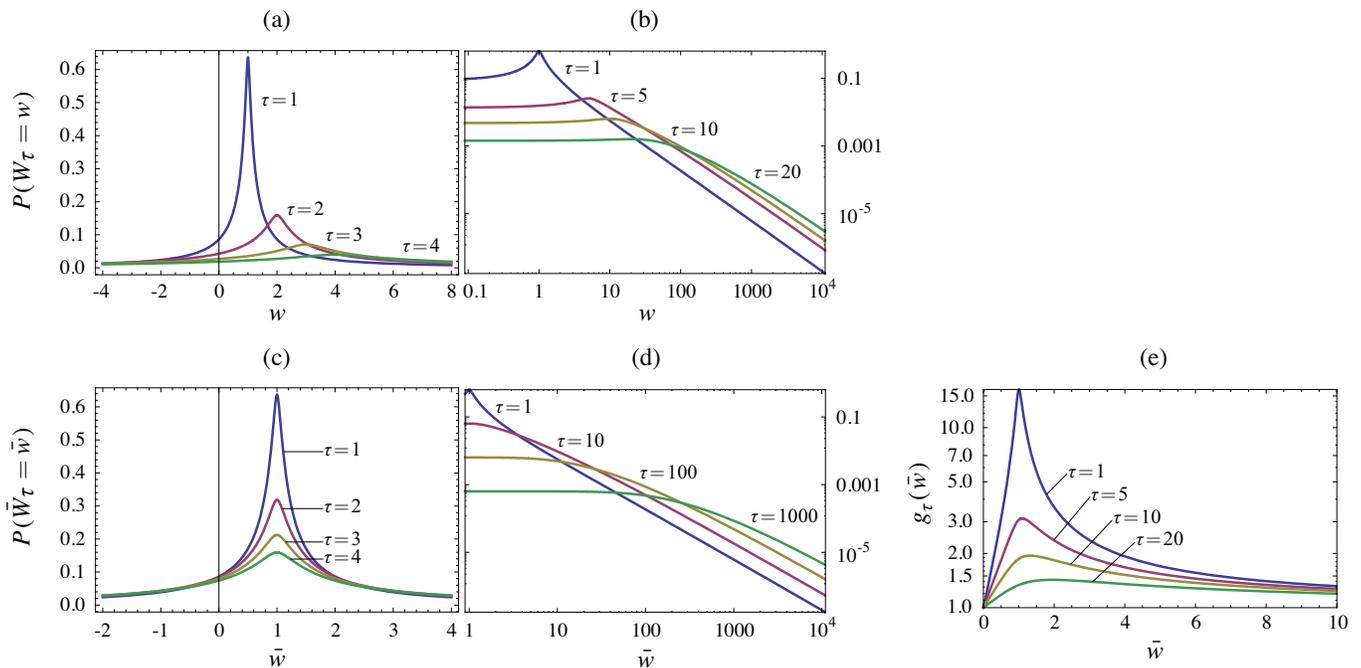}}
\caption{Work fluctuations for $\mu=0.5$ (lower L\'evy regime). (a) Distribution $P(W_\tau=w)$ of the total (extensive) work $W_\tau$. (b) Log-log plot of $P(W_\tau=w)$ for positive $w$. (c) Distribution $P(\bW_\tau=\bw)$ of the scaled work $\bW_\tau$. (d) Log-log plot of $P(\bW_\tau=\bw)$ for positive $\bw$. The power-law tails of $P(W_\tau)$ and $P(\bW_\tau)$ give rise to straight lines with slope $-\mu-1$ in the log-log plots. (e) Fluctuation relation: Log-linear plot of $g_\tau(\bw)=P(\bW_\tau=\bw)/P(\bW_\tau=-\bw)$. Units: $\tau_r=v=b=1$.}
\label{figa05}
\end{figure*}

The results of the numerical inverse Fourier transform of $G_{W_\tau}(q)$ and $G_{\bW_\tau}(k)$ are shown in Fig.~\ref{figa05} for $\mu=0.5$, which is representative of all the values in the range $(0,1)$. The distributions $P(W_\tau)$ and $P(\bW_\tau)$ that we obtain in this range of $\mu$ characterize what we call the \emph{lower L\'evy regime}, and share many of the properties that we have mentioned for the upper L\'evy regime. In particular, $\bW_\tau$ satisfies an anomalous fluctuation relation in the lower L\'evy regime, since its distribution has power-law tails and $g_\tau(\bw)\ra 1$ as $\bw\ra\infty$; see Fig.~\ref{figa05}(e).

The main difference between the lower and upper L\'evy regimes are the scaling properties of the fluctuations with $\tau$. Whereas the typical fluctuations of $\bW_\tau$ decrease with increasing $\tau$ when $\mu\in (1,2)$, they increase with $\tau$ when $\mu\in (0,1)$ because, for that interval, the typical fluctuations of the ``extensive'' work $W_\tau$ increase faster than $\tau$. As a result, $P(\bW_\tau)$ does not become more and more concentrated around its mode in the limit $\tau\ra\infty$, as was the case in the upper L\'evy regime (see Fig.~\ref{figa15}), but flattens in this limit. This implies that there is no Law of Large Numbers for $\bW_\tau$ in the lower L\'evy regime: as longer and longer integration times are considered, the typical fluctuations of the scaled work actually increase in size, which translates in Fig.~\ref{figa05} into a flattening of $g_\tau(\bw)$ with $\tau$. 

Since the typical fluctuations of $\bW_\tau$ increase with $\tau$, one might be tempted to rescale the total work $W_\tau$ by a $\tau$-factor larger than $\tau$. However, the distribution that one would obtain from this rescaling would be symmetric: it would assign the same probability to positive and negative fluctuations of equal magnitude, and would thus contain no information about the nonequilibrium steady state behavior of the model, since $g_\tau$ would then trivially be equal to 1. 

\section{Dissipated work}
\label{secdissw}

In previous studies of the model that we consider here, two quantities have been analyzed \cite{zon2004}: the first is the work done by the harmonic force on the particle, which we have just studied; the second is the heat released by the particle to its environment to maintain its nonequilibrium steady state \cite{taniguchi2007}. Since the particle's dynamics is studied in the overdamped limit, the particle has no kinetic energy, so that its total internal energy is entirely given by the potential energy, which is determined by the position $y(t)$ in the comoving frame of the potential:
\be
U(t)=\frac{(x(t)-vt)^2}{2}=\frac{y(t)^2}{2}.
\ee
In this context, the heat $Q_\tau$ was defined in \cite{zon2004} as the energy gained by the particle from the mechanical work $W_\tau$ done on it by the harmonic force over a time $\tau$ minus that part of this energy which is transformed into potential energy, that is,
\be
Q_\tau=W_\tau-\Delta U_\tau,
\label{eqec1}
\ee
where $\Delta U_\tau=U(\tau)-U(0)$ is the change in potential energy after a time $\tau$. Here we assume that $x(0)=y(0)=0$, so that $\Delta U_\tau=U(\tau)$.

It is important to note that, in the original context in which the model was studied \cite{zon2004}, the quantity $Q_\tau$ defined by Eq.~(\ref{eqec1}) was correctly interpreted as the heat because the particle is immersed in a fluid, which is responsible for both the random force and the friction force applied to the Brownian particle. Accordingly, any energy gained by the particle in the form of mechanical work that is not converted into potential energy is necessarily lost, by energy conservation, to the fluid as heat. In this sense, Eq.~(\ref{eqec1}) expresses the conservation of energy.

In our treatment of the model, the random force $\xi(t)$ is an external force, which implies that the total work done is the \textit{sum} of the work $W_\tau^\xi$ done on the Brownian particle by the random force $\xi(t)$ and the mechanical work $W_\tau$ done on the particle by the moving harmonic potential. The heat $Q_\tau$ in this case should therefore be defined, by energy conservation, as
\be
Q_\tau=W_\tau+W_\tau^\xi-\Delta U_\tau.
\label{eqec2}
\ee
This quantity $Q_\tau$ is always positive because it is the heat produced by the friction alone, and not the friction and the random force as in the case of a Brownian particle immersed in a fluid. As a result, we cannot define a fluctuation relation for this quantity. In this section, we will study therefore a different quantity having both negative and positive fluctuations. We define this quantity in analogy with Eq.~(\ref{eqec1}) by 
\be
R_\tau=W_\tau-\Delta U_\tau,
\ee
where, as before, $W_\tau$ is the mechanical work. The quantity $R_\tau$ has a clear physical interpretation: it is that part of the mechanical work $W_\tau$ that is not converted into potential energy $\Delta U_\tau$. For this reason, we call $R_\tau$ the \emph{dissipated work}. Note again that $R_\tau$ would be the heat if the only contribution to the total work done on the particle was the mechanical work $W_\tau$; see Eq.~(\ref{eqec1}). In our case, there are two distinct contributions to the total work, as shown in Eq.~(\ref{eqec2}).

The probability distribution of $W_\tau-\Delta U_\tau$ was calculated in the asymptotic limit by van Zon and Cohen \cite{zon2004} for the Gaussian noise case, $\mu=2$. The generalization of their results to L\'evy noise is not straightforward, since their calculations strongly rely on the Gaussian nature of the noise. By assuming an independence property between $W_\tau$ and $\Delta U_\tau$ for $\tau\ra\infty$, we are able, however, to obtain the tail behavior of $P(R_\tau)$, which is sufficient to determine whether the dissipated work satisfies an anomalous fluctuation relation or not. The assumption, precisely, is that $W_\tau$ and $\Delta U_\tau$ become asymptotically uncorrelated in the limit $\tau\ra\infty$, so that
\be
\lex\rme^{\rmi kR_\tau}\rex=\lex\rme^{\rmi k W_\tau}\rex\lex\rme^{-\rmi k\Delta U_\tau}\rex
\ee
in this limit. For external Gaussian noise, the distribution of $R_\tau$ obtained under this assumption can be shown, from the calculations reported by Taniguchi and Cohen \cite{taniguchi2007}, to have the same asymptotic form as the exact distribution calculated in \cite{zon2003,zon2004}. We will assume here that this independence property between $W_\tau$ and $\Delta U_\tau$ holds also for L\'evy white noise when $\tau$ becomes much larger than the relaxation time $\tau_r$.\footnote{This assumption can be argued using two basic observations. First, because the noise is white (delta-correlated), the position $y(t)$ exhibits only short time-correlations. Second, because the work $W_\tau$ is an integral of $y(t)$, it can only show a weak correlation with any single position $y(t)$ in time, and, in particular, with the last position $y(\tau)$, which determines $\Delta U_\tau$. These two arguments apply both to Gaussian and L\'evy noises.}

The characteristic function of $W_\tau$ was calculated in Sec.~\ref{secwork} for all $\mu\in (0,2]$. As for the characteristic function of the potential energy,
\be
\lex\rme^{-\rmi k\Delta U_\tau}\rex=\int_{-\infty}^{\infty}\rme^{-\rmi ky^2/2}\, P(y(\tau)=y)\, \rmd y,
\label{eqcfq1}
\ee
it has no known closed-form solution \cite{mittnik1998}. However, it is known that the distribution of $y(\tau)$ is, for large $\tau$, a L\'evy distribution with the same index $\mu$ as the noise $\xi(t)$ \cite{west1982}, so that
\be
P(y(\tau)=y)\sim c\, |y|^{-1-\mu}
\ee
as $|y|\ra\infty$, with $c$ a positive constant. Inserting this asymptotic result in the integral of Eq.~(\ref{eqcfq1}), we then obtain
\be
\lex\rme^{-\rmi k \Delta U_\tau}\rex\sim \int_{\epsilon}^\infty \rme^{-\rmi ky^2/2}\, y^{-1-\mu}\, \rmd y\sim 1-a|k|^{\mu/2}
\ee
as $k\ra 0$. In these expressions, $a$ and $\epsilon$ are positive constants. The exact value of $\epsilon$ is irrelevant for the last asymptotic result to hold, since only the tail behavior of $P(y(\tau))$ affects the scaling of the integral. Moreover, the reason for having the limit $k\ra 0$ in the asymptotics of the characteristic function is because the behavior of the tails of power-law distributions is determined only by the behavior of their characteristic functions around $k=0$, and vice versa \cite{uchaikin1999}. Combining this asymptotic result with the expression of the characteristic function of $W_\tau$ given in Eq.~(\ref{eqcfw2}), we then find
\be
\lex\rme^{\rmi kR_\tau}\rex\sim (1+\rmi Mk-V|k|^\mu)(1-a|k|^{\mu/2})
\ee
as $k\ra 0$. Since $\mu\in (0,2)$, the dominant contribution in the above expression is
\be
\lex\rme^{\rmi kR_\tau}\rex\sim 1-a |k|^{\mu/2},
\ee 
which implies that
\be
P(R_\tau=r)\sim c'\, |r|^{-1-\mu/2},\quad |r|\ra\infty,
\ee
where $c'$ is some positive constant. Thus the tails of $P(R_\tau)$ have a power-law decay, as in the case of the work, which implies that the fluctuations of the dissipated work satisfy an anomalous fluctuation relation similar to the one found for $W_\tau$. The only difference with $W_\tau$ is that the decay exponent of the tails of $P(R_\tau)$ is $\mu/2$ instead of $\mu$; see Eq.~(\ref{eqpw1}). 

This result should be contrasted with the Gaussian case, for which the fluctuations of $R_\tau$, or the heat $Q_\tau$ in that case, are known to satisfy an extended fluctuation relation \cite{zon2003,zon2004}, rather than a conventional fluctuation relation, satisfied by the work $W_\tau$. For L\'evy noise, the fluctuations of $W_\tau$ are not so different from the fluctuations of $R_\tau$, since both quantities have distributions with power-law tails. In spite of this difference, there is one property of $R_\tau$ that remains the same for both Gaussian and L\'evy noises, namely, that the large fluctuations of $R_\tau$ are mostly the result of the large fluctuations of the potential energy $\Delta U_\tau$. This property, also observed for Gaussian noise \cite{zon2003,zon2004}, can be understood here by noting that the asymptotic behavior of the characteristic function of $R_\tau$ around the origin $k=0$, which determines the behavior of the tails of $P(R_\tau)$, is determined entirely by the asymptotics of the characteristic function of $\Delta U_\tau$ around $k=0$. Thus, although $R_\tau$ has different fluctuation properties depending on the noise used (Gaussian or L\'evy), its fluctuations are mostly the result of the fluctuations of $\Delta U_\tau$ no matter what noise is applied.

\section{Fluctuation relations and large deviations}
\label{secfrldt}

It should be clear from the previous results that what differentiates conventional and extended fluctuation relations, on the one hand, and anomalous fluctuation relations, on the other hand, is the existence of a large deviation principle for the distribution of the observable studied. In the case of the scaled work $\bW_\tau$, for example, $P(\bW_\tau)$ satisfies a large deviation principle for Gaussian noise, as reported in Eq.~(\ref{eqldw1}), but not for \emph{strict} L\'evy noise with $\mu\in (0,2)$. In the case of Cauchy noise, in particular, the distribution $P(\bW_\tau)$ is such that the limit
\be
I(\bw)=\lim_{\tau\ra\infty} -\frac{1}{\tau}\ln P(\bW_\tau=\bw)
\ee
yields $I(\bw)=0$ for all $\bw$, which means in effect that $P(\bW_\tau)$ does not satisfy a large deviation principle. The same result applies to the distribution of $\bW_\tau$ obtained for all $\mu\in (0,2)$, as well as for the distribution of the dissipated work $R_\tau$ obtained for all $\mu\in (0,2)$, because all these distributions have power-law tails in this range of $\mu$. This result is also general insofar as any observable $A_\tau$ that does not obey a large deviation principle does not obey a conventional or an extended fluctuation relation, as is obvious from the definition of these two types of fluctuation relations given in Eqs.~(\ref{eqfr1}) and (\ref{eqfr2}). Whether or not $A_\tau$ satisfies in this case an anomalous fluctuation relation depends on the explicit form of $P(A_\tau)$, but we know for sure that $A_\tau$ satisfies neither a conventional nor an extended fluctuation relation.

This last observation brings us to the question of whether there exist fluctuation relations other than conventional, extended, and anomalous. In other words, if an observable $A_\tau$ does not satisfy a conventional or an extended fluctuation relation, does it necessarily satisfy an anomalous fluctuation relation? 

It is difficult at this point to answer this question, since there is nothing in principle that prevents one from imagining noises that lead to distributions that are not exactly L\'evy and yet do not obey a large deviation principle. However, the fact that there exists a link between fluctuation relations and large deviation theory restricts somehow what can be imagined. Indeed, it is known from this theory that random variables obeying large deviation principles have, in most cases, finite moments at all order. Therefore, it is natural to expect that noises having finite moments should lead to conventional or extended fluctuation relations, because they should lead to distributions having a large deviation form. This is the case for Gaussian noise, as we have seen here, but also for other noises having finite moments, including Poisson noise; see \cite{baule2009}. On the other hand, noises having one or more infinite moments should lead to anomalous fluctuation relations, since they cannot lead in general to distributions having a large deviation principle. This is the case for L\'evy noise, and it should be the case, too, for any noises having distributions with power-law tails. From this point of view, the results that we have obtained for L\'evy noise should be representative of a larger class of fluctuations arising from any noises having one or more infinite moments.

\section{Discussion}
\label{secdisc}

We close this paper with some remarks about possible experimental verifications of our results, some technical issues about L\'evy noise, and extensions of our results to nonlinear models and other power-law noises. 

(1) The fact that L\'evy noise has an infinite noise power implies, as we have mentioned in Sec.~\ref{secmodel}, that one cannot define a fluctuation-dissipation relation between the friction force in the model considered here and the power of the noise. But, as argued in that section, this is not a problem insofar as the noise is \emph{external}, i.e., that it is produced and imposed externally by a physical process which is different from the physical process giving rise to the friction. Any experimental verification of our results will have to include a noise source that has this property of being \emph{external}. Perhaps the easiest way to produce such a noise is to generate L\'evy white noise on a computer (see the remark (4) below), and to feed the output to an appropriate transducer (electrical or mechanical), which will transform the digital noise produced by the computer into a physical noise, e.g., an electric noise or a mechanical noise. 

(2) We have proposed in our previous paper \cite{touchette2007} two experiments that could be used to test our results. These experiments follow the suggestion of the previous paragraph that the noise must be external. 

In the first experiment, a solid object is placed on a solid table that is vibrated horizontally by a mechanical transducer controlled by a computer; see \cite{buguin2006}. By vibrating the table with Gaussian noise or by L\'evy noise, one should be able, in principle, to generate a steady-state motion of the object, and to study the fluctuations of work- and heat-related quantities, such as the work done on the object by the gravitational force in the case where the table is slightly tilted. An experiment of this sort is under development.\footnote{M. K. Chaudhury, private communication, 2009.} The requirement that the friction and the noise are uncoupled is obviously satisfied in this case, since the friction is the solid-solid friction, which has nothing to do with the arbitrary noise externally imposed by the vibrating table.

The second experiment that we proposed in \cite{touchette2007} is based on granular gases maintained in steady states by vibrating their container. Many studies have looked at the properties of these gases when they are vibrated by periodic forcing and by Gaussian noise (see, e.g., \cite{aumaitre2001,feitosa2004,puglisi2005,visco2005,visco2006}). A natural variation of these experiments, in view of our work, is to change the forcing signal by L\'evy white noise. This can be done, in principle, in experiments, and certainly in numerical simulations of granular gases. The quantity that is usually studied for these systems is the power injected by the vibrating force. As for the mechanical work studied in this paper, one could study the power injected, and verify that the fluctuations of the latter quantity satisfy an anomalous fluctuation relation for L\'evy noise. 

(3) A different experiment, which could be used to probe the L\'evy fluctuations of $R_\tau$, can be imagined using the analogy that exists between the fluctuations of the Brownian particle studied in this paper and the current fluctuations of small RC electrical circuits \cite{zon2004a,garnier2005}. The noise in such circuits is \emph{internal}, since it is the thermal Johnson-Nyquist noise, usually treated as Gaussian white noise. However, there is nothing that prevents one from introducing an additional, \emph{external} noise source in these circuits via fluctuating voltage sources, such as those studied, e.g., in \cite{luchinsky1998}. The fluctuations of the voltage sources can be generated by a computer, which means that they can be used, in principle, to mimic L\'evy white noise. Under Gaussian (thermal Johnson-Nyquist) noise, it is known that the fluctuations of the heat dissipated by the resistance in RC circuits follow an extended fluctuation relation \cite{zon2004a,garnier2005}. Based on our results, we expect that, in the presence of an additional L\'evy noise, the heat fluctuations will be similar to those of the dissipated work $R_\tau$ studied here.

\begin{figure}[t]
\includegraphics{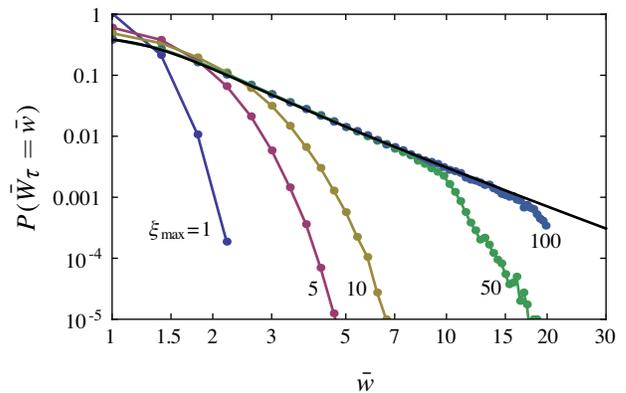}
\caption{Convergence of distributions for truncated L\'evy noise. (Colored lines) Log-log plot of $P(\bW_\tau)$ for truncated Cauchy noise with various cut-off values $\xi_{\rm max}$, which is the maximum value imposed to the noise $\xi(t)$. The curves are obtained by a direct numerical sampling of the trajectories of the Langevin equation (\protect\ref{eqle1}) with truncated Cauchy noise. (Black line) $P(\bW_\tau)$ obtained with exact Cauchy noise, as calculated in Eq.~(\protect\ref{eqpbwa1}). Units: $\tau_r=v=b=1$. Simulation parameters: $\Delta t=10^{-2}$ sec, $\tau =5$ sec.}
\label{figconv1}
\end{figure}

(4) For the purpose of the experiments just described, L\'evy noise can be generated physically to a good degree of accuracy by generating a L\'evy noise on a computer, using techniques similar to those used for generating Gaussian noise (see, e.g., \cite{samoradnitsky2000,chambers1976,janicki1993}), and by feeding this artificial noise into a mechanical or electrical transducer. Of course, L\'evy noise generated on a computer is never exactly ``L\'evy'': like any physical noise, there is always a maximum value of the noise that can be applied to a system. In the case of L\'evy noise, this maximum value or \emph{cut-off} transforms the noise into a \emph{truncated L\'evy noise} \cite{mantegna1994a} for which the distributions of $W_\tau$ and $R_\tau$ do not have the exact form of a L\'evy distribution, simply because truncated L\'evy noise has finite moments to all order. However, it is possible to reveal the L\'evy character of these distributions by studying their behavior or ``trend'' as the cut-off is increased. 

Figure~\ref{figconv1} shows, as an illustration, the distribution of $\bW_\tau$ obtained by direct sampling of Eq.~(\ref{eqle1}) using truncated Cauchy noise with different cut-off values. The behavior of $P(\bW_\tau)$ seen in this figure is general for truncated L\'evy noise \cite{mantegna1994a,koponen1995}: i.e., as the cut-off is increased, the distribution of $P(\bW_\tau)$ approaches the L\'evy distribution predicted for exact L\'evy noise, which in this case is the Cauchy distribution given in Eq.~(\ref{eqpbwa1}). The same behavior is expected to arise for $R_\tau$. 

(5) For strict L\'evy noise with $\mu\in (0,2)$, there is always a very large fluctuation that dominates the other fluctuations in time, especially in the lower L\'evy regime, $\mu\in (0,1)$. For Gaussian noise, on the other hand, all observable fluctuations are more or less of the same order of magnitude. This difference between L\'evy and Gaussian noise should directly be observable in the experiments mentioned above.

(6) The generalization of our results to nonlinear Langevin equations with L\'evy noise should give rise to interesting results. It is known, for example, that the stationary distribution of a Langevin equation involving, as here, a quadratic potential is unimodal for Gaussian and L\'evy noise. In the case of a quartic potential, however, the stationary distribution of the position is still unimodal for Gaussian noise but bimodal for L\'evy noise \cite{chechkin2002}. An interesting question, in the context of this result, is whether the distributions of $W_\tau$ and $R_\tau$ obtained for a moving quartic potential are also bimodal. For the quartic potential, it is also known that the distribution of the position has a finite variance, so it would be interesting to see whether or not the fluctuations of $W_\tau$ and $R_\tau$ are anomalous in this case.

(7) A recent paper by Chechkin and Klages \cite{chechkin2009} has appeared recently, which studies an anomalous fluctuation relation similar to the one studied in this paper and our previous work \cite{touchette2007}. The paper by Chechkin and Klages deals with L\'evy noise, as in this paper, but also with long-time correlated Gaussian noise, which gives rise to a probability distribution of the work having a stretched-exponential form \cite{chechkin2009}.

\begin{acknowledgments}
This work was supported by RCUK (Interdisciplinary Academic Fellowship), the London Mathematical Society, and the National Science Foundation, under award PHY-0501315. The hospitality of the Rockefeller University, where part of this work was carried out, is gratefully acknowledged.
\end{acknowledgments}

\bibliography{levyfluct}
\end{document}